\documentclass[
twocolumn,
]{ceurart}

\usepackage{balance}
\usepackage{caption}

\begin{document}

\copyrightyear{2021}
\copyrightclause{Copyright for this paper by its authors.
  Use permitted under Creative Commons License Attribution 4.0
  International (CC BY 4.0).}

\conference{GraphiCon 2021: 31th International Conference on Computer Graphics and Vision,
  September 27--30, 2021, Nizhny Novgorod, Russia}

\title{Objective video quality metrics application to video codecs comparisons: choosing the best for subjective quality estimation}

\author[1]{Anastasia Antsiferova}[%
email=aantsiferova.graphics.cs.msu.ru
]
\address[1]{Lomonosov Moscow State University,
  119991, Russia, Moscow, Leninskiye Gory, 1}

\author[1]{Alexander Yakovenko}[%
email=alexander.yakovenko@graphics.cs.msu.ru
]

\author[1]{Nickolay Safonov}[%
email=nikolay.safonov@graphics.cs.msu.ru
]

\author[1,2]{Dmitriy Kulikov}[%
email=dkulikov@graphics.cs.msu.ru
]
\address[2]{Dubna State University,
  141982, Russia, Moscow region, Universitetskaya, 19}

\author[1]{Alexander Gushin}[%
email=alexander.gushchin@graphics.cs.msu.ru
]

\author[1]{Dmitriy Vatolin}[%
email=dmitriy@graphics.cs.msu.ru
]

\begin{abstract}
Quality assessment plays a key role in creating and comparing video compression algorithms. Despite the development of a large number of new methods for assessing quality, generally accepted and well-known codecs comparisons mainly use the classical methods like PSNR, SSIM and new method VMAF. These methods can be calculated following different rules: they can use different frame-by-frame averaging techniques or different summation of color components. In this paper, a fundamental comparison of various versions of generally accepted metrics is carried out to find the most relevant and recommended versions of video quality metrics to be used in codecs comparisons. For comparison, we used a set of videos encoded with video codecs of different standards, and visual quality scores collected for the resulting set of streams since 2018 until 2021.\end{abstract}

\begin{keywords}
video quality rating \sep
comparison of metrics \sep
video codec comparison \sep
psnr \sep
ssim \sep
vmaf
\end{keywords}

\maketitle

\section{Introduction}

According to Cisco forecast \cite{cisco2017}, by 2022 79\% of the world's Internet traffic is expected to be video traffic. To reduce the cost of video storage and the load on data transmission channels, video compression algorithms are being actively developed and improved. Measuring of video quality plays a key role in this area. The number of studies, publications and grants allocated for the creation of new quality metrics is growing every year. One of the reasons for the lack of new metrics usage is the irreproducibility of their accuracy on large datasets and real data, therefore, generally accepted comparisons of compression algorithms still use classical methods: PSNR, SSIM, as well as the VMAF metric that has recently gained popularity. For example, to demonstrate the effectiveness of the new coding standards, the committees conduct objective testing using PSNR and subjective comparison on several \cite{boyce2018jvet} videos. Companies involved in the development of new video codecs, as well as their clients, started using VMAF, which had shown a high correlation with visual quality in many studies \cite{7986143}.

All of the above methods are commonly used for additional quality control, for example, due to its high computation speed PSNR is used in the early stages of development, when thousands of configurations need to be tested. SSIM and VMAF are tracked during intermediate stages, since they take longer to compute. The generally accepted comparisons usually demonstrate all these metrics: for example, in the comparisons conducted by Jan Ozer and published on the streaming media resource \cite{ozer_2020}, or in annual comparisons conducted in Moscow State University \cite{msuvideocodecscomparisons}. However, these generally accepted metrics have many configurations that significantly affect the ranking of the compared compression methods. For example, the PSNR or SSIM calculation can be performed only for the luminance Y-channel, or the summation of the Y, U and V channels can be performed. When summing, different coefficients can also be used (luminance is usually given a larger coefficient than other channels). Currently, there are no clear recommendations for summing and using certain channels to calculate metrics, so the interpretation of the comparison results becomes much more complicated: instead of monitoring three metrics, it may be necessary to check more than 20 indicators.

Many implementations of modern compression standards have special modes for increasing the scores of popular metrics. For example, x264, x265 video encoders have configuration modes for PSNR, SSIM. The libaom encoder has a tuning mode for VMAF. This is an open information about tuning, while in fact many commercial solutions may contain hidden settings to increase the values of generally accepted metrics, which may lead to a decrease in visual quality. During the one of subjective comparisons conducted at Moscow State University, it was shown that the --tune ssim setting improves the visual quality of the video as well as SSIM scores. However, the video preprocessing techniques used in the libaom codec's --tune vmaf setting can lead to a significant reduction in the visual quality, which was shown in \cite{siniukov2021hacking}.

In this paper, we analyzed the correspondence of various configurations of PSNR, SSIM and VMAF to the visual quality. We collected and used a special dataset for analysis, as open datasets with available visual quality ratings contained distortions from only one or two codecs (usually H.264 and H.265 standards). Therefore, special attention in the study was paid to the creation of a special set of videos encoded by various implementations of several compression standards. In this way, videos with representative types of distortions caused by video coding were obtained.
The creation of the dataset was carried out from 2018 to 2021 within subjective comparisons of video codecs conducted annualy in Moscow State University \cite{msuvideocodecscomparisons}.

\section{Related work}

There is a number of studies comparing different video quality metrics. At the same time, the correlation of each of the metrics with subjective estimates can vary greatly depending on the dataset and the type of distortion for which it was calculated. For example, the goal of \cite{4106565} was to show that metrics aimed at evaluating TV signals with high resolution, bitrate and high FPS perform much worse on video with low bitrate, resolution and variable FPS. The authors confirmed this conclusion and showed that the NTIA videoconference model \cite{nla.cat-vn4167959} method has the best accuracy, followed by the NTIA general model \cite{nla.cat-vn4167959}, Watson's DVQ \cite{Xiao_dct-basedvideo} and VSSIM \cite{WANG2004121}.

In \cite{VRANJES20131} the authors compared nine different metrics on three datasets: LIVE, ECVQ and EVVQ. ECVQ and EVVQ were created using JVT JM v.10.2 encoder (based on H.264 / AVC) and XviD v.1.1.0 (open-source encoder based on MPEG-4 Part 2 specification). The authors came to the conclusion that MOVIE and FMSE metrics perform better than others for assessing all impairments, except for simulating transmission over an IP network. In \cite{8901796}, the authors compared various metrics using 20 FullHD sequences taken from popular streaming services. Using x265 v2.7 encoder, they compressed these sequences to 10 quality levels with gradually increasing bitrate and resolution. Five metrics were used for assessing video quality on the generated videos. It was concluded that the metrics correlate much better in SD range rather than in HD range, which has noticeably fewer compression artifacts. As a result of the comparison, VMAF showed better correlation than PSNR, SSIM, MS-SSIM and VIF.

Thus, the problem of finding a suitable quality metric that is maximally correlated with visual assessment is currently important. In the existing works on assessing the correlation of video quality metrics, only a small number of video codecs have been investigated (mainly open implementations of the H.264 and H.265 standards). Therefore, the task of assessing the relevance of metrics on a wide variety of video datasets and with a large number of types of distortions remains relevant. There are many datasets for comparing the performance of algorithms, the most popular of which are LIVE-VQA \cite{5404314} and LIVE-VQC \cite{8463581}. The biggest drawback of these sets for finding best metrics for compression quality measurement is a small list of compression artifacts. Also, many new metrics that leverage machine learning have been trained on these datasets. These factors call into question the applicability of such datasets for an objective comparison of metrics performance and make it relevant to conduct a study based on an independent dataset with a representative spectrum of distortions arising during video compression.

\section{Data collection for video quality accessing algorithms analysis}

To analyze video quality metrics relevance for video codecs comparisons, a special data set was collected, including video sequences and subjective scores. Subjective comparisons were performed independently with different video sets and encoder. In each of the comparisons video set consisted of FullHD video sequences with different spatial and temporal complexity as it affects compression quality and performance. The videos were selected from over 18,000 open source clips with high bitrate downloaded after analyzing over 5 million source videos from the Vimeo website. When choosing a video set for comparison, clustering in terms of space-time complexity was used. The video selection methodology is described in \cite{zvezdakova2020bsq}.

To obtain a representative set of coding artifacts, video codecs of different standards were used: eleven H.265/HEVC encoders, five AV1 encoders, two H.264/AVC encoders and four encoders of other standards (VVC, VP9, SIF, xvc). Each video was encoded with three different target bitrates: 1000, 2000 and 4000 Kbps. This range was chosen to simplify the subjective comparison procedure, since at higher bitrates the video quality is more difficult to distinguish visually.

Subjective assessment was carried out via pairwise comparisons using Subjectify.us platform. The Bradley-Terry model was used to transform the results of pairwise voting into scoring for video streams. A detailed description of the methodology is presented on the service website. To increase the relevance of the results, at least 10 responses were received from participants for each pair of encoded video streams. The number of subjective ratings per pair depended on the confidence intervals: more responses were received for complex or similar videos which were hard to be distinguished.

Table~\ref{tab:datastat} presents a summary information about created datasets.
\begin{table*}
  \caption{Sizes of datasets used to analyze video quality metrics}
  \label{tab:datastat}
  \begin{tabular}{ccccc}
    \toprule
    Dataset & Number of codecs & Number of test videos & Number of encoded streams & Number of responses\\
    \midrule
    \textbf{CC-2018}&10&5&150&22542 \\
    \textbf{CC-2019}&11&5&165&25784 \\
    \textbf{CC-2020}&11&8&264&236736 \\
    \textbf{UGC-2020}&7&10&210&35232 \\
    \textbf{Total}&39&28&789&320294 \\
  \bottomrule
\end{tabular}
\end{table*}

For all encoded videos, various configurations of PSNR, SSIM, MS-SSIM, VMAF and NIQE objective metrics were calculated. The following versions of the PSNR algorithm were considered:

\begin{itemize}
\item PSNR avg. MSE - in this version, when aggregating frame-by-frame scores for the entire video, first the arithmetic mean for the MSE is calculated, and then the logarithm is taken.
\end{itemize}
\begin{displaymath}
  \operatorname{PSNR}_{\text {avg.} MSE}(V, \hat{V})=10 \log _{10} \frac{MAX_{I}^{2}}{\frac{1}{n} \sum_{i=1}^{n} MSE\left(V_{(i)}, \hat{V}_{(i)}\right)}
\end{displaymath}
\begin{itemize}
\item PSNR avg. log - when aggregating frame-by-frame scores for the entire video, the PSNR is calculated for each frame, and then the arithmetic mean value of all video frames is calculated.
\end{itemize}
\begin{displaymath}
\operatorname{PSNR}_{avg . \log}(V, \hat{V})=\frac{1}{n} \sum_{i=1}^{n} 10 \log _{10} \frac{MAX_{I}^{2}}{M S E\left(V_{(i)}, \hat{V}_{(i)}\right)}
\end{displaymath}

Different versions of the VMAF metric were also considered:
\begin{itemize}
\item VMAF 0.6.1, VMAF 0.6.2, VMAF 0.6.3
\item VMAF 0.6.1 NEG  - version called ``no enhancement gain'', less sensitive to artificial increasing by preprocessing
\item For the above four models, the Phone version was additionally calculated, and a version for 4K video for the VMAF 0.6.1 model was calculated.
\end{itemize}

For all reference methods, the following calculation options for color components were considered: Y - metrics are calculated only for the luminance channel, YUV4:1:1 - metrics are calculated independently for three components and the result is averaged as 4*Y+U+V, and similar methods for YUV6:1:1, YUV8:1:1, YUV10:1:1, and for rarely used YUV1:1:1, YUV2:1:1.
To calculate the metric values, the MSU VQMT version 12.1 \cite{msu_vqmt} was used.

\section{Comparison results}

Since separate subjective comparisons were conducted for each year video sets, it was necessary to obtain an overall correlation across the entire dataset for each metric. For each metric, the Fisher z-transform was applied to all correlations on individual sequences, and the obtained values were used to calculate the weighted mean and confidence interval with weights proportional to the number of distortions. The final correlations were calculated using the inverse transformation \cite{doi:10.1080/00221309809595548}.

\begin{figure*}
  \centering
  \includegraphics[width=\linewidth]{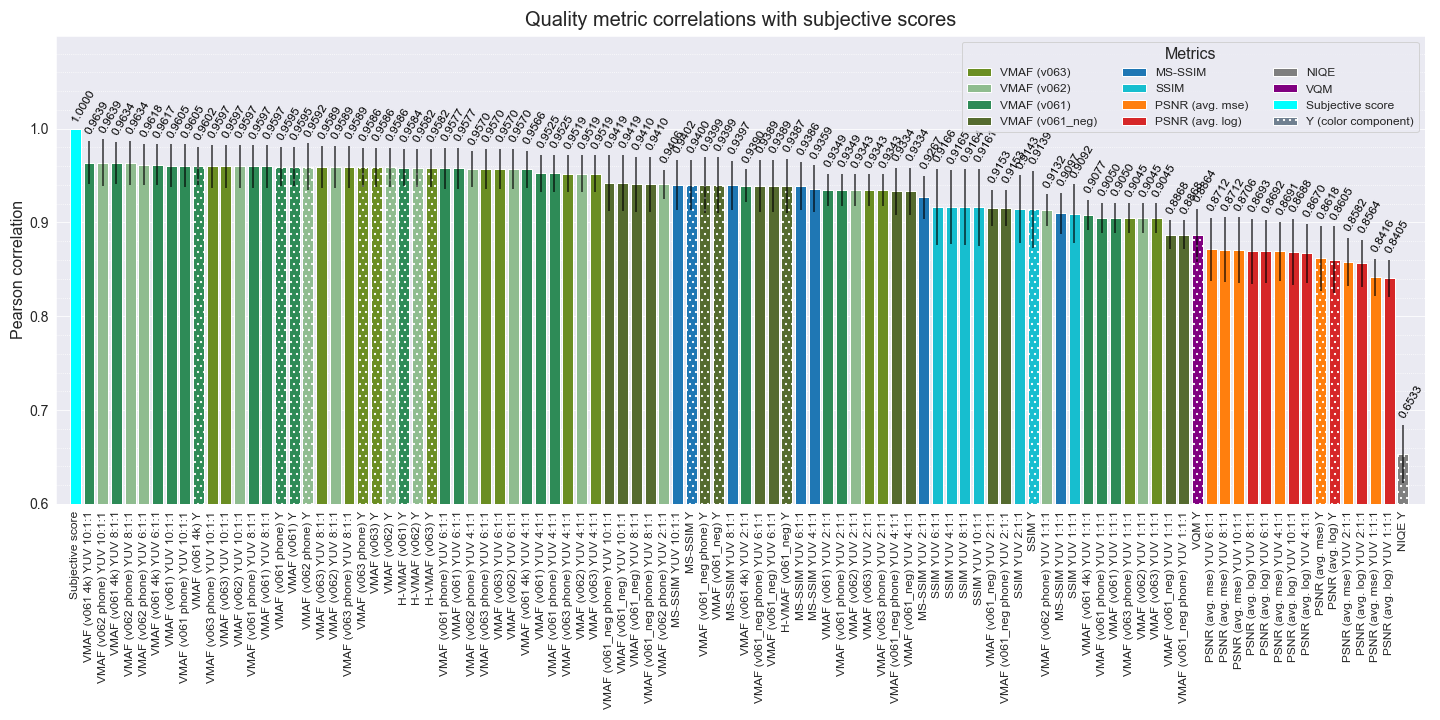}
  \caption{Pearson correlation between objective metrics scores and ranking scores received from visual assessments.}
  \label{fig:allpearson}
\end{figure*}

\begin{figure*}
  \centering
  \includegraphics[width=\linewidth]{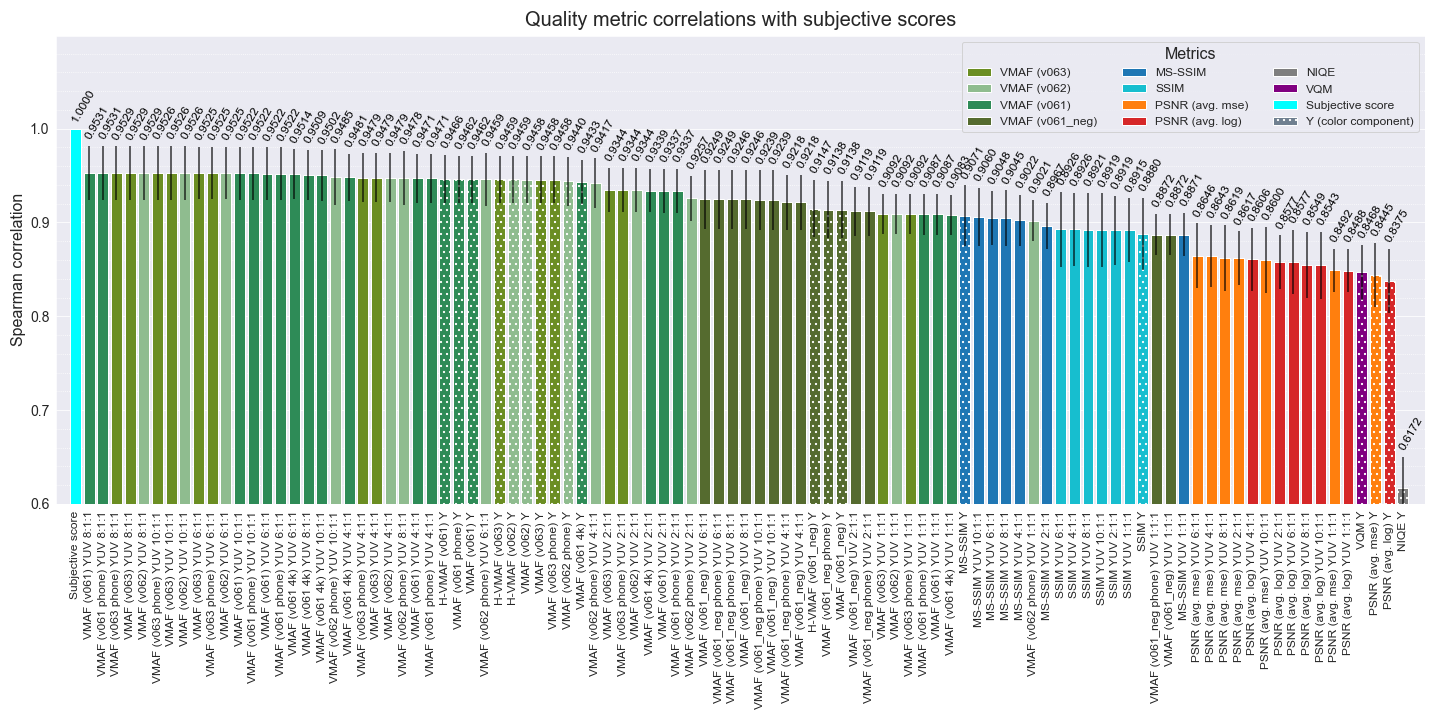}
  \caption{Spearman correlation between objective metrics scores and ranking scores received from visual assessments.}
\label{fig:allspearman}
\end{figure*}

Fig.~\ref{fig:allpearson} and Fig.\ref{fig:allspearman} show the obtained Spearman and Pearson correlations. Colors indicate different groups of metrics. The graphs show that VMAF variations have the highest correlation with subjective quality scores, and the difference between them is insignificant. They are followed by VMAF NEG and MS-SSIM, also with almost equal correlation. PSNR versions have the lowest correlation among full-reference metrics.

\subsection{Comparison of different YUV summations}

\begin{figure*}[htb]
\centering
 \begin{minipage}[b]{.65\linewidth}
   \begin{minipage}[b]{.9\textwidth}
    \centerline{\includegraphics[width=\linewidth]{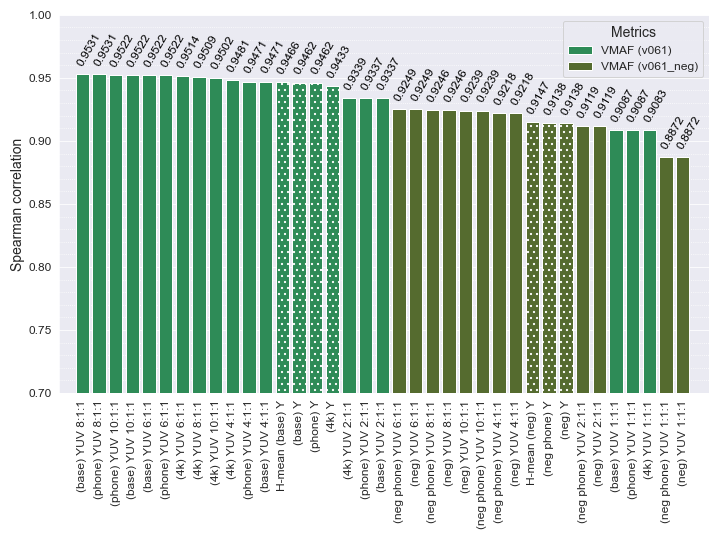}}
    \captionsetup{labelformat=empty}
    \caption*{VMAF v0.6.1 and its modifications, including VMAF NEG}
   \end{minipage}
   \begin{minipage}[b]{.45\textwidth}
    \centerline{\includegraphics[width=\linewidth]{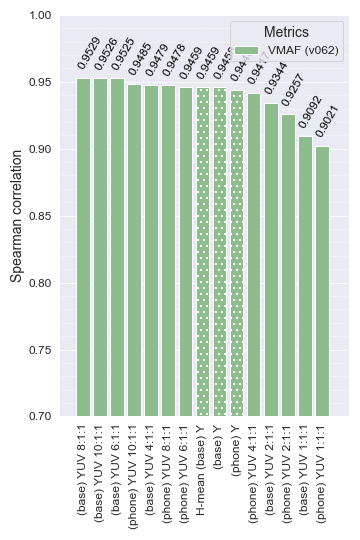}}
    \captionsetup{labelformat=empty}
    \caption*{VMAF v0.6.2 and its modifications}
   \end{minipage}
   \begin{minipage}[b]{.45\textwidth}
    \centerline{\includegraphics[width=\linewidth]{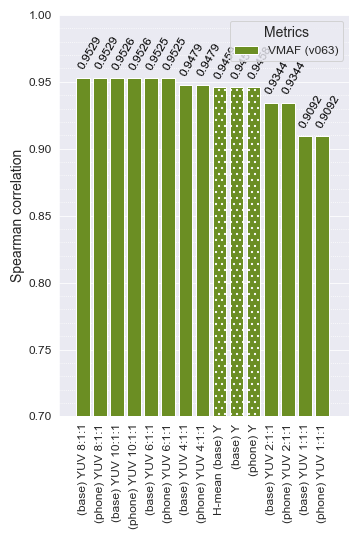}}
    \captionsetup{labelformat=empty}
    \caption*{VMAF v0.6.3 and its modifications}
   \end{minipage}
 \end{minipage}
 \begin{minipage}[b]{.33\linewidth}
   \begin{minipage}[b]{.9\textwidth}
    \centerline{\includegraphics[width=\linewidth]{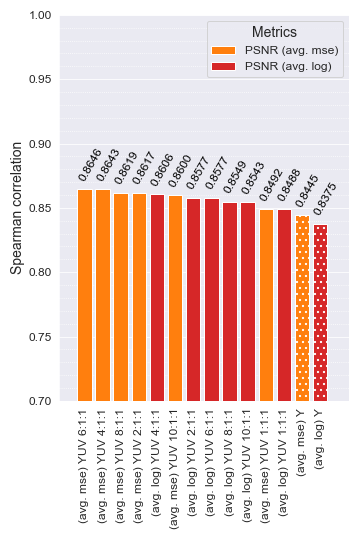}}
    \captionsetup{labelformat=empty}
    \caption*{PSNR with different frame-by-frame \\ averaging strategies}
   \end{minipage}
   \begin{minipage}[b]{.9\textwidth}
    \centerline{\includegraphics[width=\linewidth]{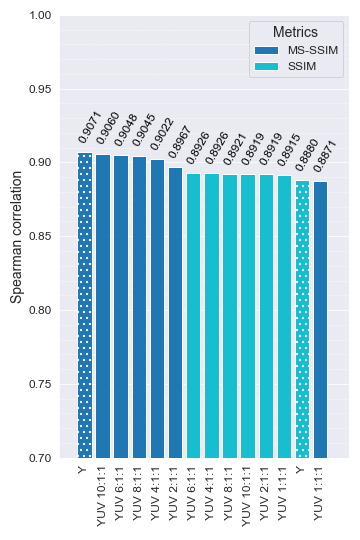}}
    \captionsetup{labelformat=empty}
    \caption*{SSIM and MS-SSIM}
   \end{minipage}
 \end{minipage}
 \caption{Performance comparison of various metrics depending on YUV summation strategy. In the notation (e.g. 4:1:1), the numbers denote the coefficients that were proportional to the weights of the metric values for the Y, U, and V color components.}
 \label{fig:lotspearman}
\end{figure*}

Fig.~\ref{fig:lotspearman} presents the results of various summations for the analyzed metrics. In all cases, YUV1:1:1 and YUV2:1:1 show worse results comparing to other ways of summing channels. Different versions of some metrics (for example, MS-SSIM) are almost identical, but there is no one uniquely best YUV summation for all metrics.

Table~\ref{tab:sum} shows the best options for summing components of different metrics. Some summation options have almost identical correlation, in the table they are separated by commas. For VMAF, the best coefficients for YUV summation are 6:1:1, 8:1:1 and 10:1:1, for PSNR 6:1:1 and 4:1:1, for SSIM. For MS-SSIM almost all summing methods showed the same accuracy.
\begin{table*}[htb]
  \caption{Variants of Y, U, V components summing that have the best correlation with visual quality for different metrics.}
  \label{tab:sum}
  \begin{tabular}{lc}
    \toprule
    Metric & Best YUV Summing\\
    \midrule
    VMAF 0.6.1 & \multirow{9}{*}{8:1:1, 10:1:1, 6:1:1}\\
    VMAF 0.6.2 \\
    VMAF 0.6.3 \\
    VMAF 0.6.1 phone \\
    VMAF 0.6.1 neg \\
    VMAF 0.6.1 neg phone\\
    VMAF 0.6.1 4K \\
    VMAF 0.6.2 phone \\
    VMAF 0.6.3 phone \\
    \midrule
    PSNR avg. MSE & \multirow{2}{*}{6:1:1, 4:1:1} \\
    PSNR avg. log \\
    \midrule
    SSIM&No significant difference between summing methods \\
    \midrule
    MS-SSIM & Y (only luma component), 10:1:1, 6:1:1, 8:1:1 \\
    
  \bottomrule
\end{tabular}
\end{table*}

\subsection{Metrics comparison for different encoding standards}

\begin{figure}
  \centering
  \includegraphics[width=\linewidth]{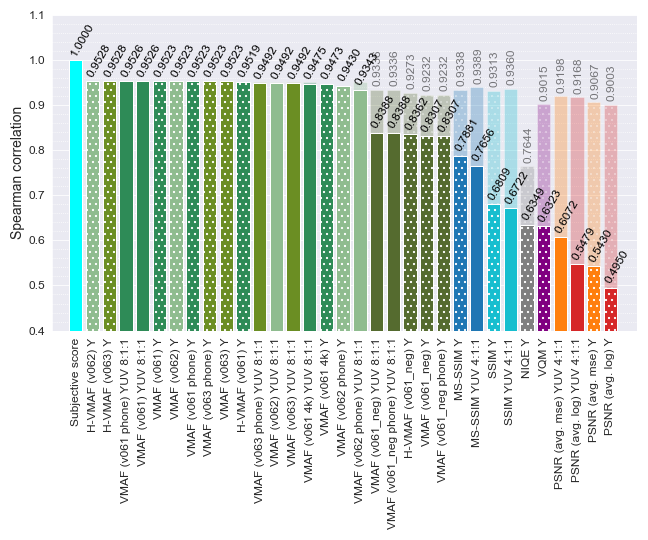}
  \caption{Comparison of metric correlations for video streams encoded by codecs of AV1 standard (saturated colors) and a set of video streams encoded with H.265 encoders (semi-transparent columns). PSNR has low relevance for analyzing the quality of AV1 video streams.}
  \label{fig:av1_vs_h265_spearman}
\end{figure}

When considering subsets of videos encoded with different standards, the results for AV1-encoded streams differs from other standards. Fig.~\ref{fig:av1_vs_h265_spearman} demonstrates the difference in metrics correlations at videos encoded with AV1 and H.265 encoders. The correlation between PSNR and visual scores is much smaller than for videos encoded with codecs of other standards. This may happen due to the use of neural networks in new generation encoders, which restore the boundaries of objects that cause pixel-by-pixel similarity violation penalized by PSNR. The best results for such video streams were shown by multiscale metrics (VMAF, MS-SSIM).

The most stable values are shown by all metrics at H.264/AVC-encoded streams. The correlations of all metrics on these videos exceed 0.94.

\subsection{Metrics comparison for different videos}

\begin{figure*}[htb]
\centering
 \begin{minipage}[b]{.45\linewidth}
  \centerline{\includegraphics[width=\linewidth]{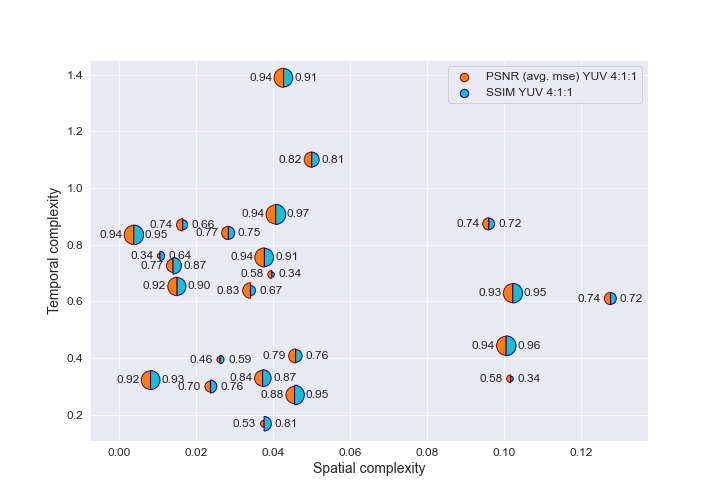}}
  \caption*{(a) PSNR and SSIM}
 \end{minipage}
 \begin{minipage}[b]{.45\linewidth}
  \centerline{\includegraphics[width=\linewidth]{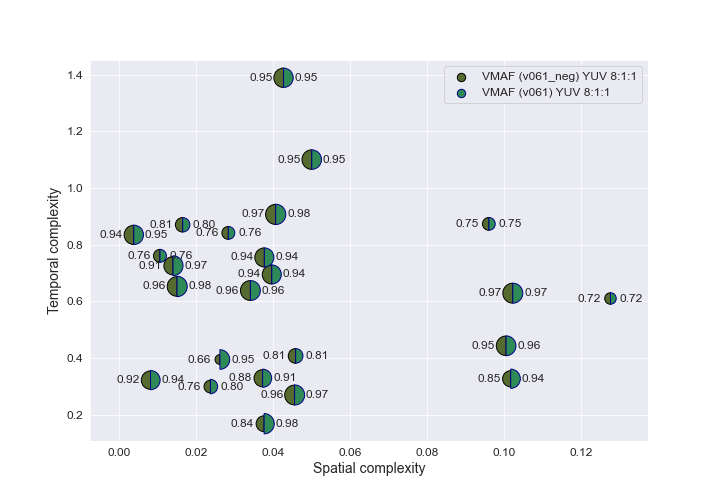}}
  \caption*{(b) VMAF and VMAF NEG}
 \end{minipage}
 \caption{Comparison of different metrics correlations on videos from our dataset in coordinates of videos spatial and temporal complexity.}
 \label{fig:siti_corr}
\end{figure*}

Fig.~\ref{fig:siti_corr} shows videos space-time complexity distribution and the correlation between PSNR and VMAF for different videos. For some videos that are simple in terms of spatial and temporal complexity, the relevance of PSNR and SSIM is very different from VMAF. The difference might occur due to other factors, so it is better to track multiple metrics to measure performance on individual videos.

\subsection{Metrics comparison for different bitrates}

\begin{figure}
  \centering
  \includegraphics[width=\linewidth]{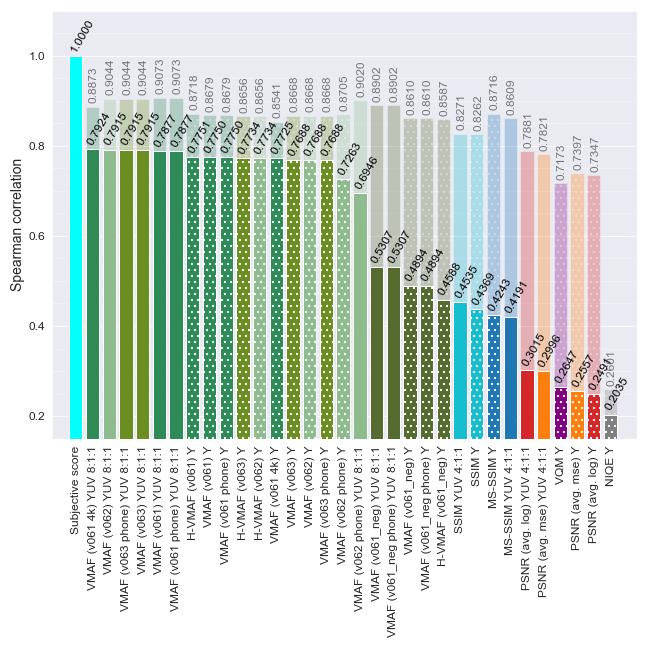}
  \caption{Comparison of metric correlations for video streams encoded with low bitrate (semi-transparent colors) and high bitrate (saturated colors). Video streams with low bitrate are easier to compare visually, and the correlation between metrics and visual scores is higher.}
  \label{fig:low_vs_high_all_spearman}
\end{figure}

Fig.~\ref{fig:low_vs_high_all_spearman} shows that the relevance of the metrics for high bitrate encoded video streams is lower than for low bitrate video streams. This may happen due to the difficulty of good quality videos visual ranking, while it is easier to compare videos containing artifacts at low bitrates. From the results we can conclude that VMAF reflects the visual perception of artifacts much better than other metrics for high bitrates.

\subsection{Comparison of VMAF and VMAF NEG}

\begin{figure}
  \centering
  \includegraphics[width=\linewidth]{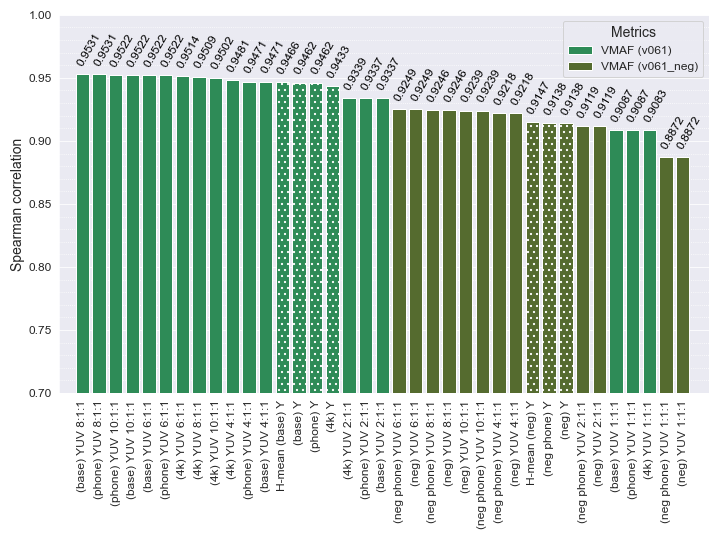}
  \caption{Comparison of VMAF (green) and VMAF NEG (dark green). VMAF NEG shows lower correlation with visual quality.}
  \label{fig:av1_vs_h265_all_spearman}
\end{figure}

VMAF NEG showed smaller correlation with visual quality than VMAF (Fig.~\ref{fig:av1_vs_h265_all_spearman}). The developers from NETFLIX recommended using VMAF NEG for video codec comparisons, as it reduces the chance of cheating and increasing of metric values through video preprocessing. However, if you are comparing different versions of the same algorithm, or it is known for certain that the algorithms do not artificially increase VMAF scores, it is better to use classical VMAF to obtain a more accurate estimate of the visual quality.

\subsection{Comparison of SSIM and MS-SSIM, PSNR avg. MSE and PSNR avg. log}

\begin{figure*}[htb]
\centering
 \begin{minipage}[b]{.45\linewidth}
  \centerline{\includegraphics[width=\linewidth]{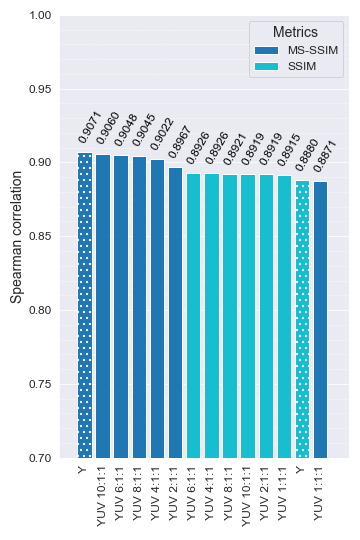}}
  \caption*{(a) SSIM versions comparison}
 \end{minipage}
 \begin{minipage}[b]{.45\linewidth}
  \centerline{\includegraphics[width=\linewidth]{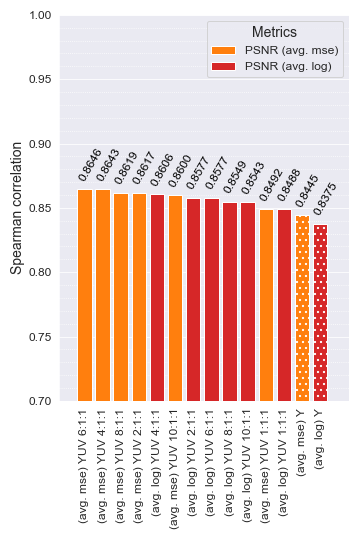}}
  \caption*{(b) PSNR versions comparison}
 \end{minipage}
 \caption{Comparison of SSIM and MS-SSIM versions, PSNR avg. MSE and PSNR avg. log. versions.}
 \label{fig:ssim_and_psnr_versions}
\end{figure*}

MS-SSIM shows higher correlation with visual quality than classical SSIM. Correlation of different PSNR versions with visual quality is less different than between SSIM and MS-SSIM, but PSNR avg. MSE is slightly more relevant than PSNR avg. log (Fig.~\ref{fig:ssim_and_psnr_versions}).

\section{Conclusion}

This article describes the results of different versions of popular objective methods comparison with subjective quality. A large number of compression algorithms were used to perform the analysis and reveal best metrics versions for the task of video codecs comparisons. A large-scale dataset of 789 encoded streams distorted by 39 video codecs of H.264, H.265, AV1, VP9 and other standards were used, as well as 3 different bitrates.
A visual comparative analysis of the obtained sequences was carried out using Subjectify.us service, several hundred viewers took part in the comparison. The analysis of a large number of various metrics modifications was performed, including various versions of these metrics (methods of averaging the values between frames, options for accounting of color and brightness components, etc.)

Analysis of the results showed the following:
\begin{enumerate}
\item {VMAF and its different versions showed higher correaltion with visual quality that other metrics. However, recent research showed that if videos were specially prepared (preprocessed) for this metric \cite{zvezdakova2019hacking, siniukov2021hacking} visual quality may significantly decrease and the correlation becomes negative. At the same time, at high bitrates, VMAF significantly outperformes the results of other metrics (correlation is 0.7 versus 0.25-0.45)}.
\item {When calculating metrics for all color planes of YUV space, different metrics have different best summation methods:
\begin{itemize}
\item{For VMAF, best results for summing over Y, U, V is a 8:1:1 ratio}
\item{For VMAF NEG 6:1:1 is better}
\item{For SSIM - 6:1:1}
\item{For PSNR (avg. MSE) - 6:1:1}
\item{For PSNR (avg. log) - 4:1:1}
\end{itemize}}
\item{MS-SSIM showed better results than SSIM}
\item{For modifications of the PSNR metric (avg. log and avg. MSE) no significant differences were found}
\item{It was also found that when analysing AV1 codecs or AV1-encoded videos, the use of any modifications of the PSNR metric is not justified}
\item{With similar averaged correlations for all videos, some metrics may show significantly lower results in some videos. Therefore, when drawing conclusions, comparing the quality of various video encoding or processing algorithms, it makes sense to use a set of metrics while taking into account their possible sharp fluctuations on individual outputs from the test set}.
\end{enumerate}

Taking into account the above results, comparing various video coding algorithms using objective quality metrics is a complex process with many issues and peculiarities. Ignoring these features may lead to results that do not have significant justification, and sometimes contradict the results of subjective visual analysis. That is why the representatives of the codec industry and the professional community generally recognize only well-known comparisons of codecs, which should be carried out in laboratories by experienced teams in collaboration with codec developers and industry representatives. These comparisons have a correct methodological basis and empirical confirmation of all the discussion decisions used in them (for example, the choice of objective quality metrics, their modifications, methods of averaging color components, etc.). Otherwise, with a certain selection of video data, metrics, and their parameters, it is often possible to obtain not an objective comparison result, but the one desired by the researcher.

\begin{acknowledgments}
  This work is partially supported by Russian Foundation for Basic Research under Grant 19-01-00785a.
  Special thanks to the team of the Graphics and Media Laboratory of Moscow State University for valuable advice and support of our projects.
\end{acknowledgments}

\balance
\bibliography{sample-ceur}

\end{document}